\documentclass[10pt]{iopart}

\usepackage{iopams}
\expandafter\let\csname equation*\endcsname\relax
\expandafter\let\csname endequation*\endcsname\relax
\usepackage{amsmath}  
\usepackage{amsfonts} 
\usepackage{verbatim}
\usepackage{color}
\usepackage[english]{babel}
\usepackage{blindtext}
\usepackage[pdftex]{graphicx}
\usepackage{epstopdf} 
\usepackage{hyperref}
\usepackage{cite}
\usepackage[version=4]{mhchem} 
\usepackage{xspace}
\usepackage[normalem]{ulem}

\begin{document}

\title{Novel Setup for Detecting Short-Range Anisotropic Corrections to Gravity}

\author{Jake S. Bobowski$^1$,  Hrishikesh Patel$^{2,3}$, Mir Faizal$^{3}$} 
\address{$^1$Irving K. Barber Faculty of Science, University of British Columbia, 3333 University Way, Kelowna, British Columbia, Canada V1V 1V7} 
\address{$^2$Department of Physics and Astronomy, University of British Columbia, 6224 Agricultural Road, Vancouver, British Columbia, Canada V6T 1Z1}
\address{$^3$Canadian Quantum Research Center, 204-3002 32nd Avenue, Vernon, British Columbia, Canada V1T 2L7}
\ead{jake.bobowski@ubc.ca}

\begin{abstract}
In this paper we argue that, even though there are strong theoretical and empirical reasons to expect a violation of spatial isotropy at short distances, contemporary setups for probing gravitational interactions at short distances have not been configured to measure such spatial anisotropies.  We propose a simple modification to the state-of-the-art torsion pendulum design and numerically demonstrate that it suppresses signals due to the large spatially-isotropic component of the gravitational force while maintaining a high sensitivity to short-range spatial anisotropies. We incorporate anisotropy using both Yukawa-type and power-law-type short-distance corrections to gravity. The proposed differential torsion pendulum is shown to be capable of making sensitive measurements of small gravitational anisotropies and the resulting anisotropic torques are largely independent of the details of the underlying short-distance modification to gravity. Thus, if there is an anisotropic modification to gravity, from any theory, in any form of the modified potential, the proposed setup  provides a practical means of detecting it. 

~


~

\noindent{(Some figures may appear in colour only in the online journal)\/}
\end{abstract}

\ioptwocol 

\section{Introduction}
Even though general relativity, and its short-distance approximation in the form of Newtonian gravity, does not break the isotropy of space, there are strong reasons to suspect that, at short distances, the general relativistic description will be modified and this could in turn break isotropy at short distances. In fact, as  general relativity is only an effective field theory approximation to some quantum theory of gravity, it is expected that the observed symmetries of spacetime might only be symmetries of this low-energy effective field theory, and there could exist high-energy sectors in the theory that break these symmetries. In several approaches to quantum gravity, such as causal sets~\cite{j2, j4} and loop quantum gravity~\cite{lq12, lq14, lq16, lq18}, geometry is discrete at the Planck scale. In such theories, an isotropic continuum is only a large-distance approximation to the underlying discrete geometry and higher-order corrections are expected to break spatial isotropy.  It is known that modifications to quantum mechanics, in the form of a deformation of the Heisenberg algebra, arise from such theories of quantum gravity~\cite{p1, p2}. Such modifications have recently been proposed to break isotropy at a detectable scale~\cite{mir}. 

In any theory of quantum gravity, which could break isotropy, the leading order term has to be the Einstein–Hilbert action. Because the short-distance approximation to the Einstein–Hilbert action leads to Newtonian gravity, which preserves isotropy, we expect any anisotropic terms to emerge from corrections to Newtonian gravity.  Motivated by the extra dimensions that arise in string theory, brane world models that produce either a Yukawa~\cite{17} or power-law~\cite{17a} correction to Newtonian gravity have been constructed~\cite{bw, corrected}.  Even though most brane world models are isotropic, it has also been possible to construct anisotropic models that produce anisotropic corrections to Newtonian gravity~\cite{b5, b6, b7, b8}. We also note that most spin-independent corrections to Newtownian gravity result in a linear combination of Yukawa-type and power-law-type corrections~\cite{av2, Lee:2020, adel4, aoki, b1, b2}.  An advantage of using brane world models is that, in such models, it is possible for the effective Planck scale to be much larger than the actual Planck scale, and this can lead to corrections to the gravitational field at sub-millimeter and micrometer length scales~\cite{zw12, zw14}. If the actual cosmology is described by anisotropic branes~\cite{b5, b6, b7, b8}, there could exist detectable anisotropic corrections to the gravitational field at short distances.

In conjunction with the aforementioned theoretical reasons, there are strong observational indicators pointing towards an anisotropy in space. Recent cosmological data indicates that there are clear anisotropies in the Cosmic Microwave Background (CMB)~\cite{1z,2z}. In fact, it has already been suggested that these anisotropies could occur due to short-distance modifications to the gravitational field~\cite{q12, q14}, which could have important consequences during  inflation~\cite{1y, 2y, infl, infla}. The CMB anisotropies~\cite{1z,2z} have also been studied using  anisotropic brane world models~\cite{b4,b9}. In these models, the anisotropies occur at short distances, and thus should be detectable by sensitive tabletop experiments.    

\section{Proposed Experiment}

Some of the most sensitive tests of the gravitational force at the sub-millimeter scale are made using precision torsion pendulums placed in close proximity to a rotating attractor~\cite{Lee:2020, Tan:2020}.  These tests are, in part, designed to look for deviations from the Newtonian inverse-square law (ISL).  For example, the ``missing mass'' torsion pendulum and attractor of the E\"ot-Wash group each have $N$ equally-spaced holes a radial distance $R$ from the rotation axis.  This geometry results in a periodic torque that completes $N$ cycles with each rotation of the attractor~\cite{a1v}.  Strong constraints on the Yukawa corrections to Newtonian gravity have also been extracted from measurements of the Casimir force between sinusoidally-corrugated surfaces~\cite{Bezerra:2010, Klimchitskaya:2013, Klimchitskaya:2017}.  The torsion pendulum and Casimir force methods are sensitive to distinct ranges of the Yukawa parameters and are, therefore, complementary measurements~\cite{Lee:2020, Klimchitskaya:2017}.

\begin{figure}
    \centering
    \includegraphics[width=0.75\columnwidth]{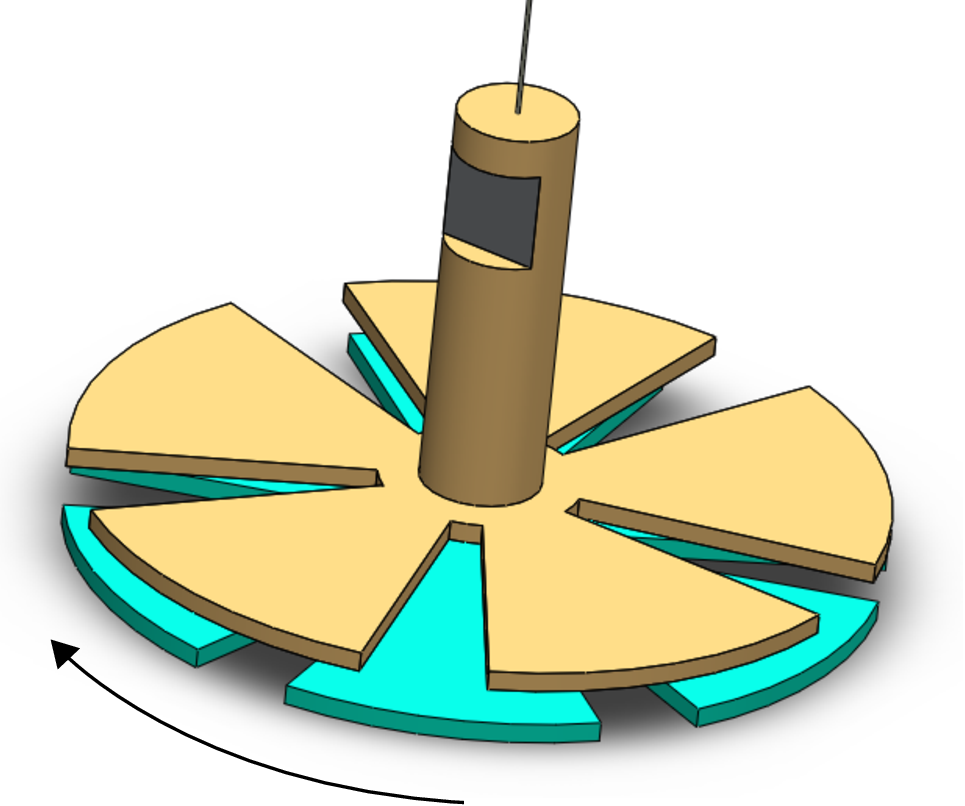}
    \caption{Diagram of the proposed experimental setup to test for short-range gravitational anisotropies.  In this example, a pendulum (gold) with $N - 1 = 5$ fins is suspended above an attractor (teal) with $N = 6$ fins.  The attractor rotates at a constant angular velocity beneath the pendulum.}
    \label{fig:pendulum}
\end{figure}

\begin{figure}
    \centering
    \begin{tabular}{c}
    (a)\includegraphics[width = 0.7\columnwidth]{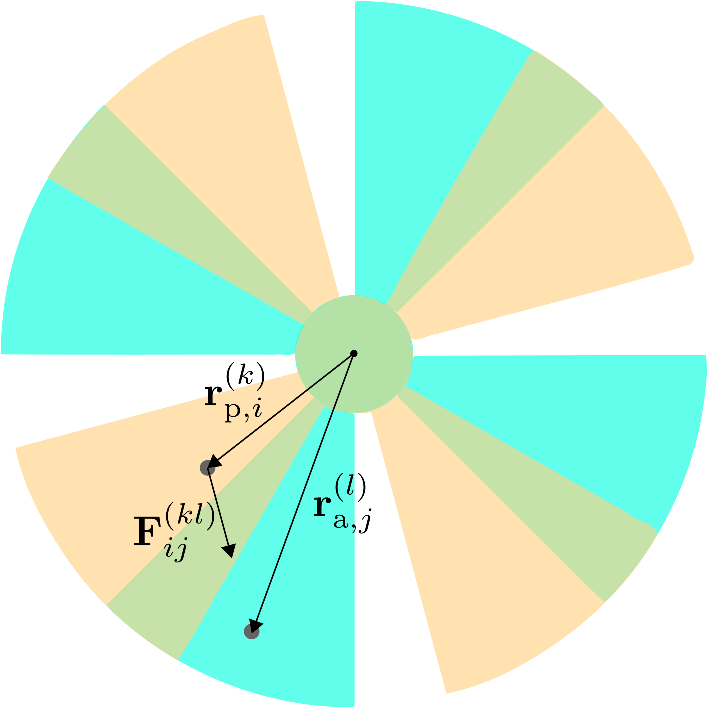}\\
    ~\\
    (b)\includegraphics[width = 0.7\columnwidth]{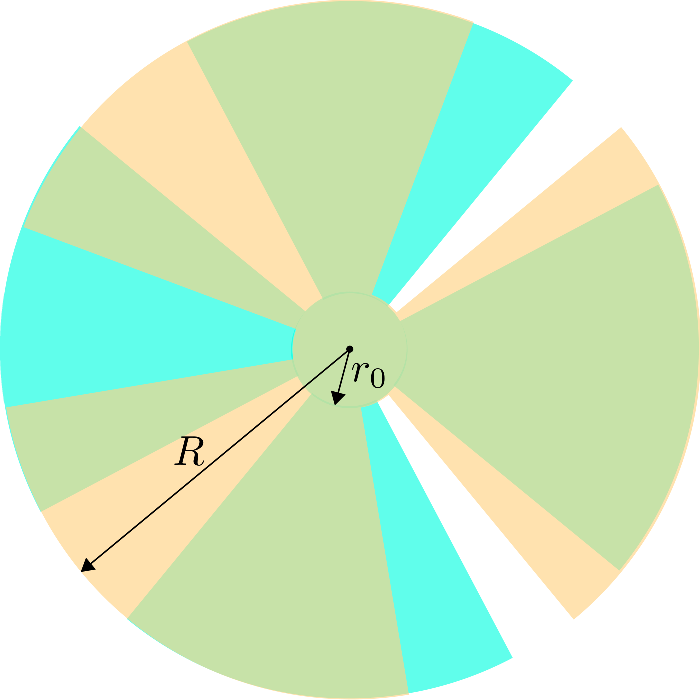}
    \end{tabular}
    \caption{(a) A symmetric torsion pendulum in which the suspended pendulum and attractor have the same number of fins.  The vector $\mathbf{r}_{\mathrm{p}, i}^{\left(k\right)}$ points from the rotation axis to the center-of-mass position of patch $k$ in pendulum fin $i$, while $\mathbf{r}_{\mathrm{a},j}^{\left(l\right)}$ points to patch $l$ in attractor fin $j$.  (b) A differential torsion pendulum with $N$ attractor fins and $N-1$ pendulum fins.  This apparatus is designed to be sensitive to gravitational anisotropies.}
    \label{fig:pend_attract}
\end{figure}

Our objective was to design a modified pendulum-attractor set that is insensitive to spatially-isotropic gravitational forces while maintaining a high sensitivity to a posited spatially-anisotropic component of the gravitational force.  Our differential torsion pendulum design, shown in Fig.~\ref{fig:pendulum}, consists of an attractor with $N$ fins and a pendulum with $N - 1$ fins.
Figure~\ref{fig:pend_attract} shows a top view of the pendulum (gold) and attractor (teal) fins.  Semi-transparent coloring has been used to show the overlap of pendulum and attractor fins.  Figure~\ref{fig:pend_attract}(a) shows the traditional setup in which the pendulum and attractor have the same number of fins.  In this configuration, which is sensitive to the isotropic component of the gravitational force, the net torque on the pendulum vanishes only when the pendulum and attractor fins are perfectly aligned (i.e.\ the attractor is rotated by an integer multiple of $2\pi/N$ with respect to the pendulum) or perfectly misaligned (the attractor is rotated by an odd multiple of $\pi/N$ with respect to the pendulum).   

Figure~\ref{fig:pend_attract}(b), on the other hand, shows an attractor with $N$ fins and a pendulum with $N - 1$ fins.  In this case, to within a first order approximation, the net torque on the pendulum due to the isotropic component of the gravitational force vanishes for a large number of attractor orientations.  As discussed below, the suppression of the isotropic component of the gravitational torque is enhanced as $N$ is increased.  Note that the angular extent of the fins differ in Figs.~\ref{fig:pend_attract}(a) and (b).  In (a), the fins occupy half of the available area and in (b) they occupy $75\%$ of the available area.  The reason for using a larger fin size with the differential pendulum is discussed in Sec.~\ref{sec:anisotropic}.

\begin{figure}
    \centering
	\includegraphics[width = 0.9\columnwidth]{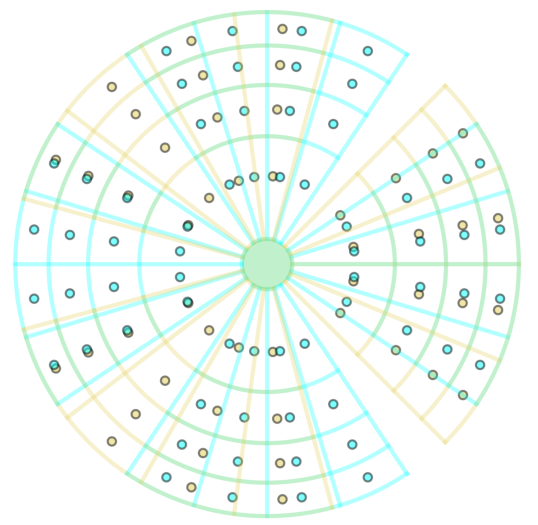}
    \caption{$N - 1 = 3$ pendulum fins (gold) and $N = 4$ attractor fins (teal) each subdivided into a $4\times 4$ grid of patches.  The patches of the pendulum fins each have mass $m_\mathrm{p}$ and the attractor patches have mass $m_\mathrm{a}\ne m_\mathrm{p}$.  The straight lines along the radial direction and the arc segments define the patch boundaries.  The points indicate the centre-of-mass positions of each patch.}
    \label{fig:coordPlot}
\end{figure}

To calculate the net torque on the pendulum, the pendulum and attractor fins were first subdivided into ``patches'' of equal mass $m_\mathrm{p}$ and $m_\mathrm{a}$, respectively.  Each patch was then approximated as a point mass located at the center-of-mass position of the patch.  Figure~\ref{fig:coordPlot} reproduces the geometry of Fig.~\ref{fig:pend_attract}(b) with $N - 1 = 3$ pendulum fins and $N = 4$ attractor fins, but highlights the patch boundaries and center-of-mass positions.  For clarity, only a small number of patches ($4\times 4$) were used in the figure. 
In the absence of Yukawa or power-law corrections, the force on point mass $k$ in pendulum fin $i$ due to point mass $l$ of attractor fin $j$ is given by 
\begin{equation}
\mathbf{F}_{ij}^{\left(kl\right)} = \frac{G m_\mathrm{p}m_\mathrm{a}}{\left(r_{ij}^{\left(kl\right)}\right)^2}\,\hat{r}_{ij}^{\left(kl\right)},
\end{equation}
where \mbox{$\mathbf{r}_{ij}^{\left(kl\right)} = \mathbf{r}_{\mathrm{a},j}^{\left(l\right)} - \mathbf{r}_{\mathrm{p},i}^{\left(k\right)}$} is a vector that points from the center of mass of patch $k$ in pendulum fin $i$ to the center of mass of patch $l$ in attractor fin $j$, and $\hat{r}_{ij}^{\left(kl\right)}$ is a unit vector in the direction of $\mathbf{r}_{ij}^{\left(kl\right)}$.  Figure~\ref{fig:pend_attract}(a) shows the position vectors used to construct $\mathbf{r}_{ij}^{\left(kl\right)}$.  Finally, the net torque on the pendulum about the vertical axis is given by the $z$-component of $\displaystyle{\sum_{ij} \sum_k\sum_l \mathbf{r}_{\mathrm{p},i}^{\left(k\right)}\times \mathbf{F}_{ij}^{\left(kl\right)}}$.

\section{Isotropic Gravity}\label{sec:isotropic}

\begin{figure}
    \centering
    \begin{tabular}{c}
    (a)~\quad\includegraphics[width=0.85\columnwidth]{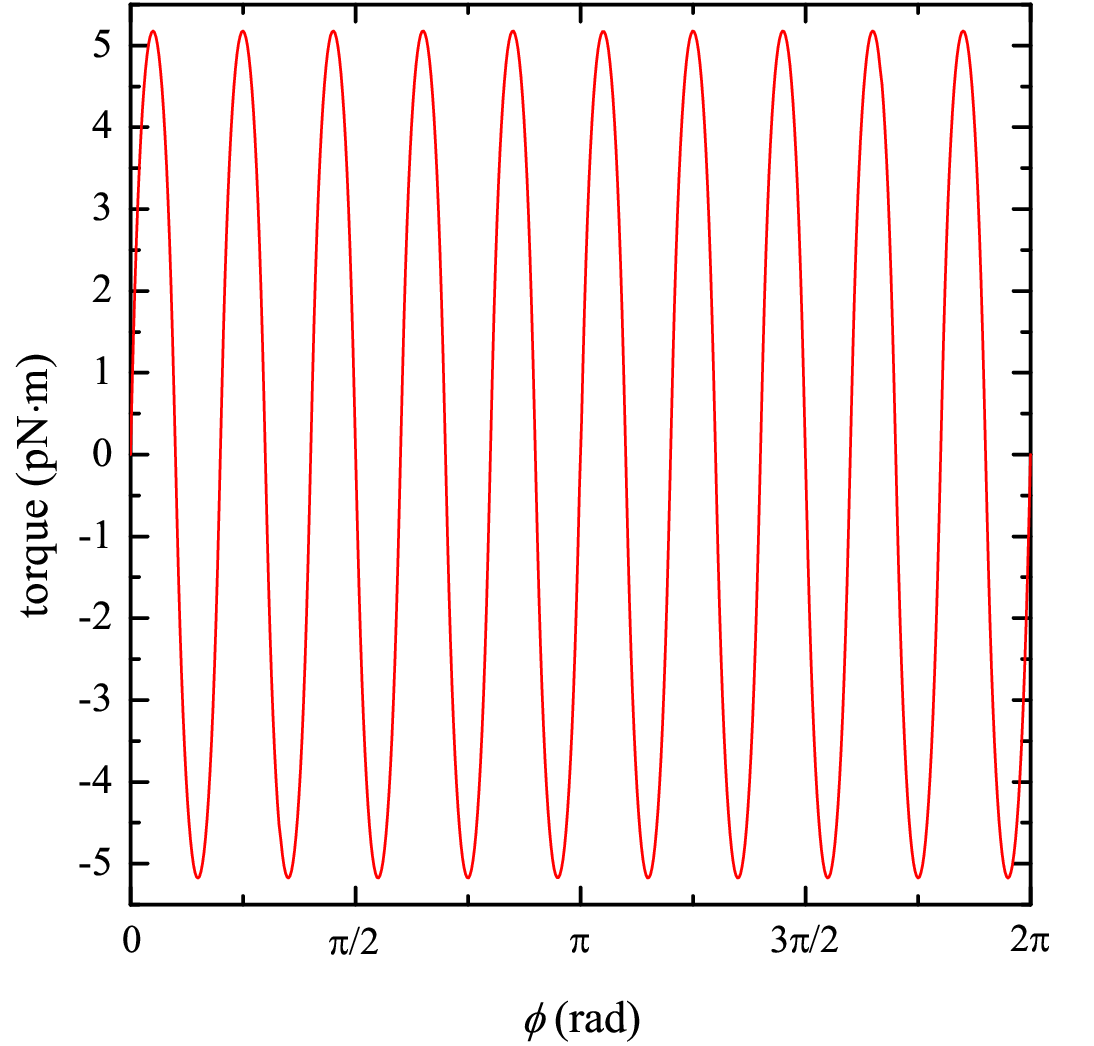}\\
    ~\\
	(b)\includegraphics[width=0.805\columnwidth]{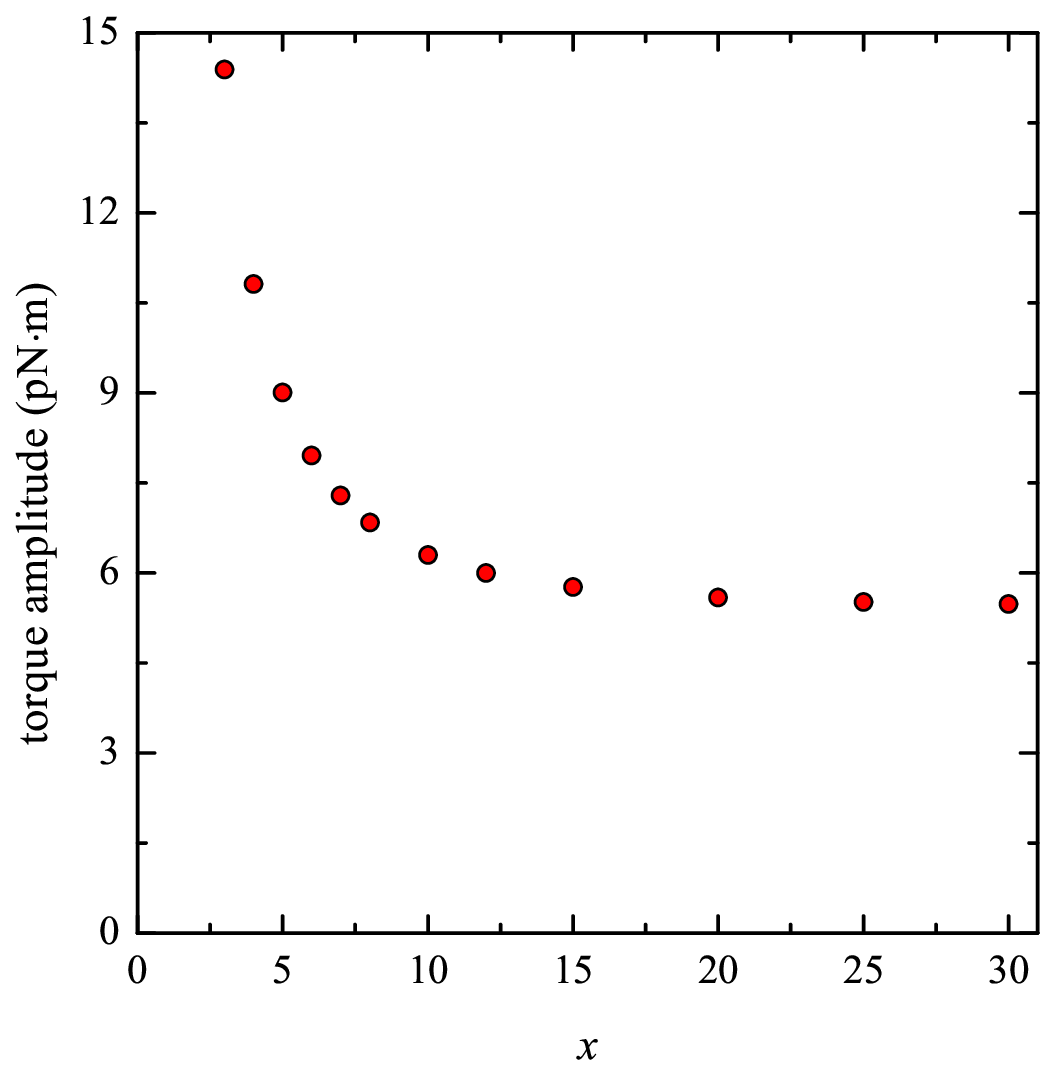}  
	\end{tabular}  
    \caption{The calculated torque as a function of attractor orientation $\phi$ for a symmetric torsion pendulum with $N = 10$ fins.  For this calculation, each pendulum and attractor fin was subdivided into $x\times x$ patches, where $x = 30$.  The torque is periodic in $\phi$ with the period $2\pi/N$.  (b) Amplitude of the torque oscillation at the fundamental frequency as a function of $x$.}
    \label{fig:torque}
\end{figure}

For verification purposes, the net torque on a pendulum with $N = 10$ fins suspended above an attractor with $N = 10$ fins was calculated as a function of the attractor's angular position $\phi$ assuming a spatially-isotropic gravitational force.  For this calculation, and all subsequent calculations, the pendulum and attractor fins were assumed have inner and outer radii of $r_0 = 5\rm\ mm$ and $R = 50\rm\ mm$, respectively.  See Fig.~\ref{fig:pend_attract}(b) for the definitions of $r_0$ and $R$.  Additionally, the fins were given a thickness $t = 1\rm\ mm$ and assumed to be made of tungsten (density $\rho = 19.3\rm\ g/cm^3$).  Finally, the bottom of the pendulum was assumed to be a distance $d = 50\rm\ \mu m$ above the top of the attractor. Figure~\ref{fig:torque} shows the results of the calculation when each pendulum and attractor fin is subdivided into $x\times x$ patches, where $x = 30$.  For each fin, there were $x + 1$ patch boundaries along the radial direction and $x + 1$ boundaries along the angular direction.
The calculated torque has the expected periodicity of $2\pi/N$ and is consistent with an independent calculation of the net torque using the numerical methods described in Ref.~\cite{Lee:2021}.

For the $50\rm\ \mu m$ separation distance, the torque for $x = 30$ exhibits a nearly pure sinusoidal oscillation.  Evaluating the Fourier transform of the data in Fig.~\ref{fig:torque}(a) reveals that the first non-zero harmonic     in angular reciprocal space occurs at $3N/\left(2\pi\right)$ and has a peak height that is more than an order of magnitude less than the height of the fundamental peak at $N/\left(2\pi\right)$.  

In order to validate the results presented above, we investigated how the calculated torque changes as $x$ is varied.  Figure~\ref{fig:torque}(b) shows the oscillation amplitude at the fundamental frequency as a function of $x$ for \mbox{$3\le x\le 30$}.  Initially, the torque amplitude drops rapidly as $x$ increases before asymptotically approaching a stable value.  From $x = 20$ to $30$, the amplitude of the torque oscillation decreased by only $2\%$.  

\section{Anisotropic Gravity}\label{sec:anisotropic}
To investigate the effects that a short-range, spatially-anisotropic component of the gravitational force would have on our differential torsion pendulum, we considered both Yukawa-like and power-law-like corrections to Newtonian gravity.  The gravitational anisotropy was applied only to the correction terms added to conventional Newtonian gravity which allowed us to gauge the universality of such anisotropies when applied to different gravitational models. As the Einstein–Hilbert action preserves isotropy, only the correction terms can break the isotropy of space. 

The anisotropic Yukawa-corrected force of point mass $l$ in attractor fin $j$ on point mass $k$ in pendulum fin $i$ is expressed as 
\begin{multline}
\mathbf{F}^{\mathrm{a},\left(kl\right)}_{\mathrm{Y},ij} = \frac{Gm_\mathrm{p} m_\mathrm{a}}{\left(r_{ij}^{\left(kl\right)}\right)^2}\Bigg[\hat{r}_{ij}^{\left(kl\right)} + \\ \alpha \left(1 + \frac{r_{ij}^{\left(kl\right)}}{\lambda}\right)e^{-r_{ij}^{\left(kl\right)}/\lambda}\,\hat{r}^{\mathrm{a},\left(kl\right)}_{ij}(\varepsilon) \Bigg].\label{eq:FY}
\end{multline}
Here, $\alpha$ is the strength correction parameter and $\lambda$ is the scale correction parameter.  As described in the Supplemental Material, the gravitational anisotropy is introduced by defining an anisotropic unit vector $\hat{r}^{\mathrm{a},\left(kl\right)}_{ij}(\varepsilon)$ that is characterized by an anisotropy parameter $\varepsilon\ll 1$.  In a similar way, an anisotropic power-law corrected force~\cite{17a} can be expressed as
\begin{equation}
    \mathbf{F}^{\mathrm{a},\left(kl\right)}_{\mathrm{P},ij} = \frac{G m_\mathrm{p} m_\mathrm{a}}{\left(r_{ij}^{\left(kl\right)}\right)^2} \left[ \hat{r}_{ij}^{\left(kl\right)} + \left(\frac{k}{r_{ij}^{\left(kl\right)}}\right)\,\hat{r}^{\mathrm{a},\left(kl\right)}_{ij}(\varepsilon)\right].\label{eq:FP}
\end{equation}
In this case, the correction term is parameterized by a single constant $k$. 

For the isotropic torque calculation presented in Sec.~\ref{sec:isotropic}, the pendulum and attractor fins each occupied half of the total available area $\pi(R^2 - r_0^2)$.  Figure~\ref{fig:pend_attract}(a) shows a top view of this geometry for the $N = 4$ case.  For the case of the differential torsion pendulum with $N - 1$ pendulum fins and $N$ attractor fins, we found that the desired anisotropic torque vanished when the fins occupied half of the available area.  The Supplemental Material presents a systematic study of the magnitude of the anisotropic torque as a function of the angular size of the pendulum and attractor fins.  The anisotropic torque vanishes when the fins occupy $0$, $50$ or $100\%$ of the available area and peaks when they they fill $25$ and $75\%$ of the available area.  The peak at $75\%$ filling is slightly higher, presumably because of the increased masses of the pendulum and attractor fins.  All of the numerical results  presented in this section for the differential pendulum were obtained using fins that occupy $75\%$ of the available area.  Figures~\ref{fig:pendulum}, \ref{fig:pend_attract}(b), and \ref{fig:coordPlot} all show differential pendulum geometries using the $75\%$ filling.

\begin{figure}
    \centering
    \begin{tabular}{c}
    (a)\includegraphics[width=0.85\columnwidth]{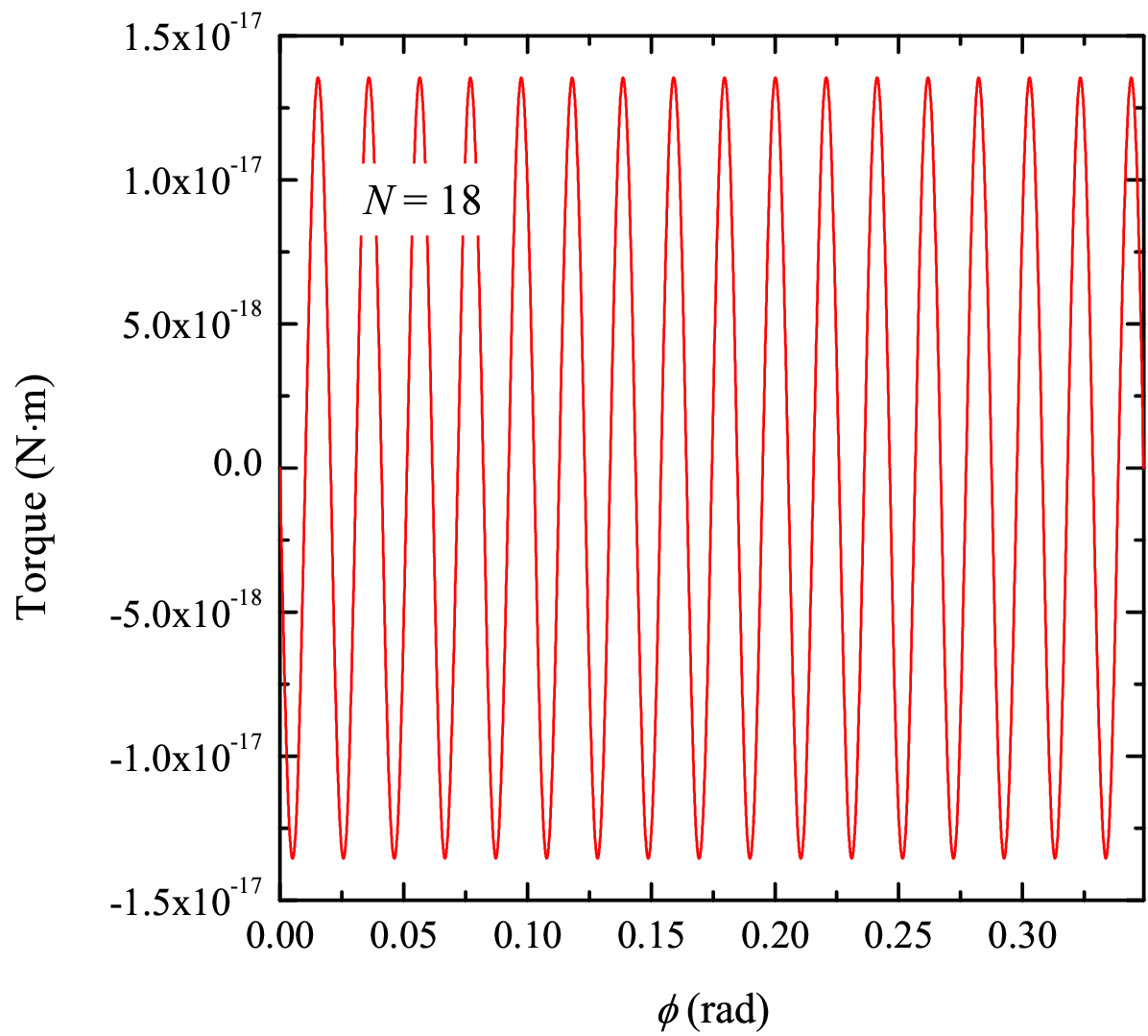}\\ (b)\quad~\includegraphics[width=0.85\columnwidth]{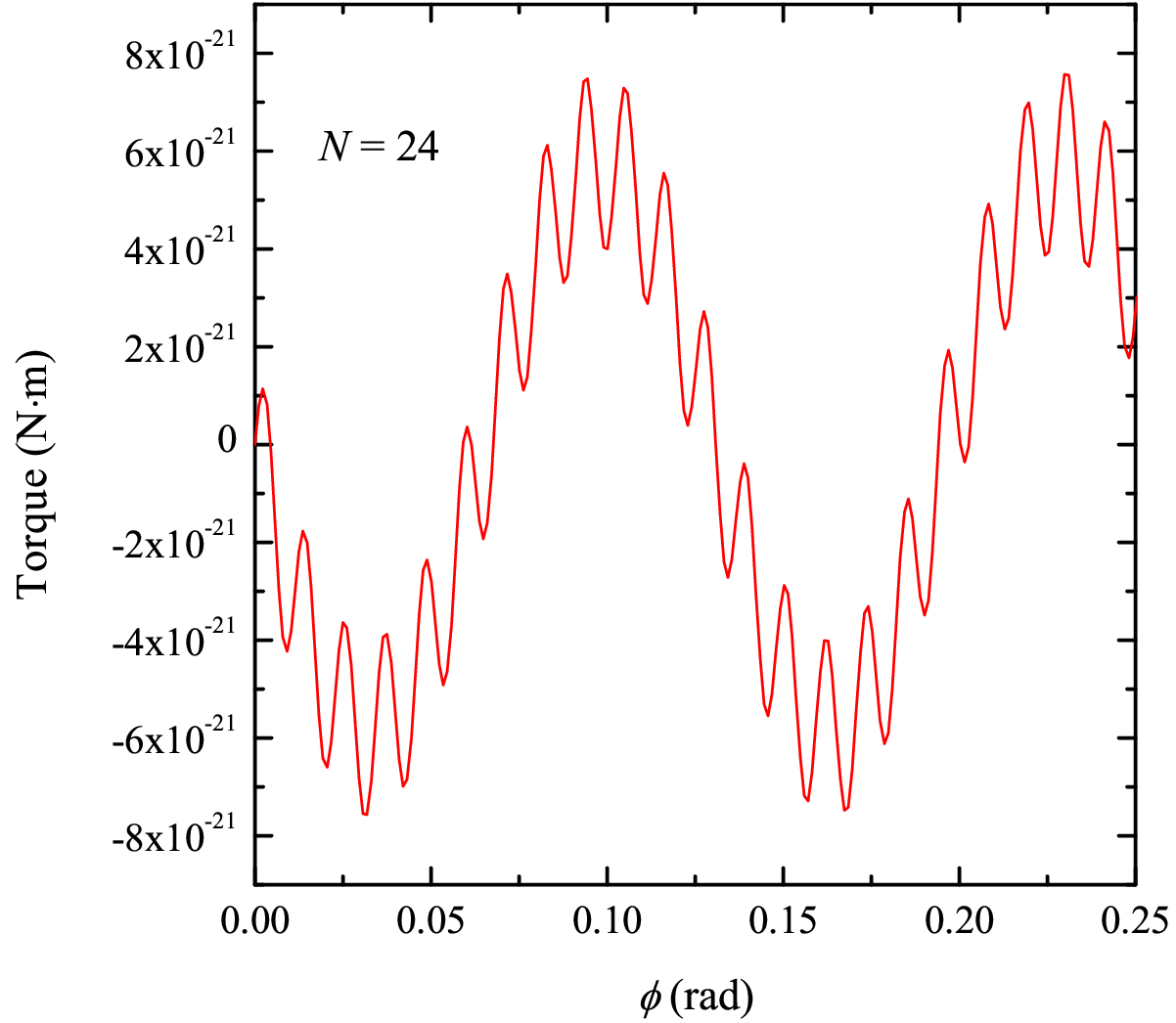} \\ (c)\quad~\includegraphics[width=0.85\columnwidth]{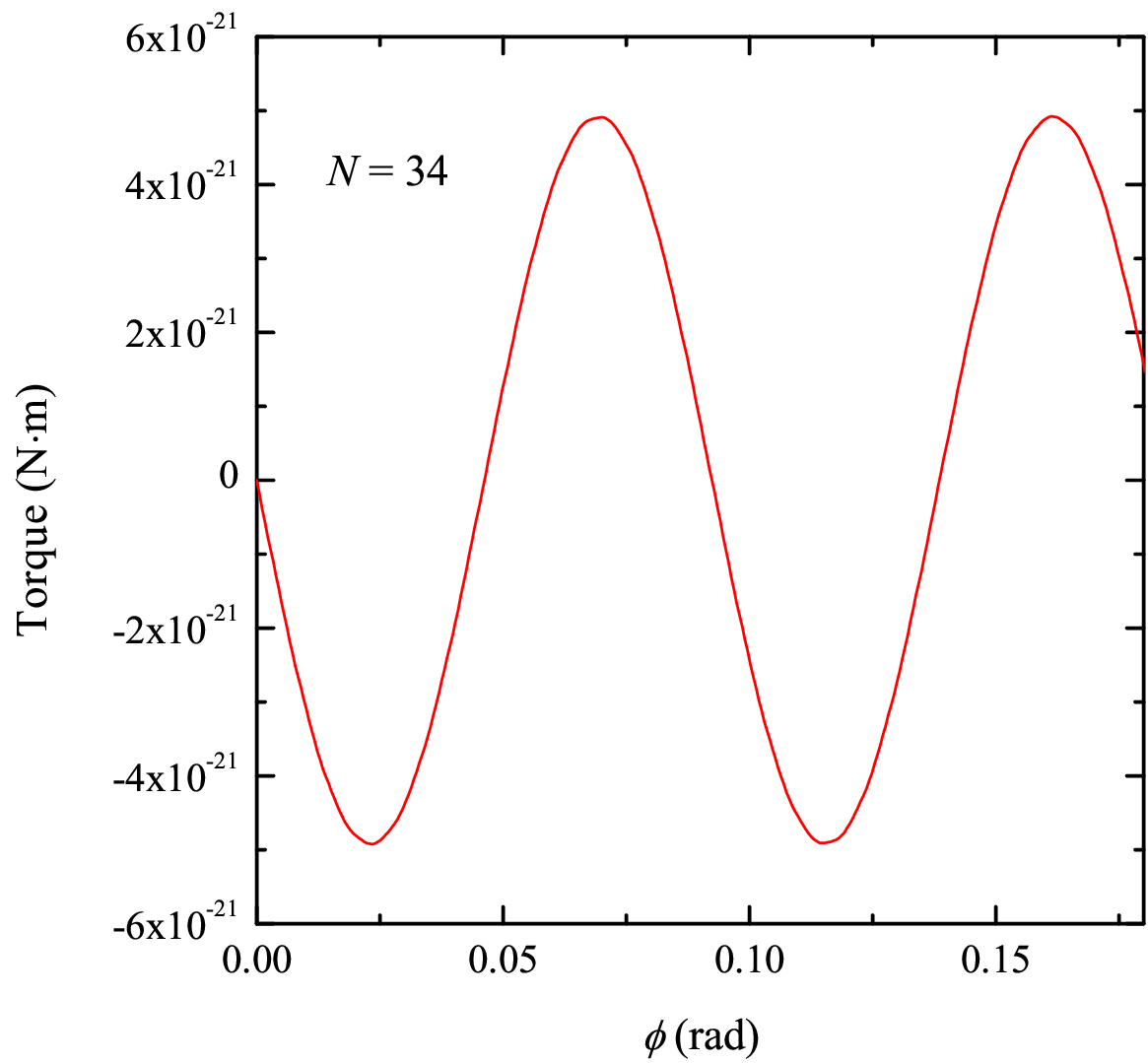}
    \end{tabular}
    \caption{Calculated torque oscillations in the presence of a Yukawa-type gravitational anisotropy for $N$ attractor fins and $N-1$ pendulum fins.  For both the attractor and pendulum, the fins occupy $75\%$ of the available area. (a)  For $N = 18$, the large isotropic component of the gravitational torque dominates. (b) For $N = 24$, the isotropic and anisotropic components of the torques are comparable.  (c) For $N = 34$, the large suppression of the isotropic component of the torque allows the anisotropic component to dominate.}
    \label{fig:torqueOsc}
\end{figure}

\begin{figure}
    \centering
    \includegraphics[width=0.85\columnwidth]{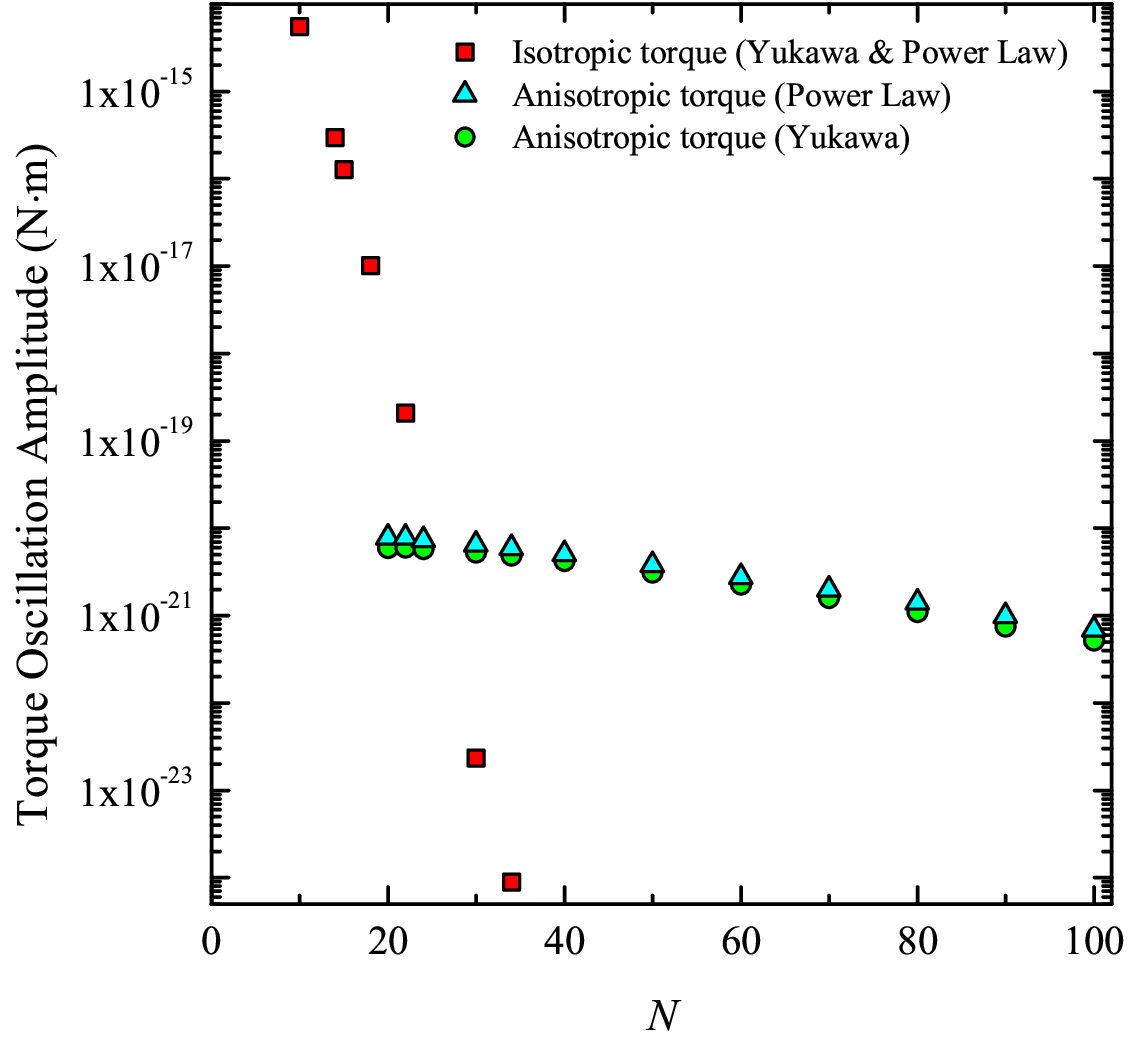}
    \caption{Amplitudes of the isotropic (squares) and anisotropic torque oscillations as a function of $N$.  The circles and triangles show the anisotropic amplitudes of the Yukawa-corrected and power-law-corrected gravitational forces, respectively.  The squares represent the isotropic amplitudes for both the Yukawa and power-law corrections.}
    \label{fig:AmpVsN}
\end{figure}

An adaptive scheme was implemented to set the sizes of the $x\times x$ patch grids used for the pendulum and attractor fins.  For small values of $N$, the large angular size of the fins requires more patches to achieve the same accuracy obtained for the large-$N$ calculations.  The strategy used was to set an upper limit for the acceptable mass of each patch.  The value of $x$ in the $x\times x$ patch grid was then incremented until the patch mass fell below this threshold.  For the torque versus $\phi$ data shown in Fig.~\ref{fig:torqueOsc}, the upper limit on the patch mass was set to $5\rm\ mg$ which, for $1$-mm thick tungsten fins, corresponds to a patch area of $0.26\rm\ mm^2$.  The torque oscillation amplitude versus $N$ data presented in Fig.~\ref{fig:AmpVsN} were less computationally expensive which allowed us to lower the upper limit on the patch mass by a factor of four to $1.25\rm\ mg$. 

Figure~\ref{fig:torqueOsc} shows the numerically-calculated torque on a pendulum with $N-1$ fins suspended above an attractor with $N$ fins as a function of the attractor's angular position $\phi$.  For each value of $N$ considered, the torque about the rotation axis of the pendulum was calculated assuming an isotropic gravitational force ($\varepsilon=0$) and a gravitational force with a short-range anisotropy ($\varepsilon=10^{-4}$) included. The calculations assumed that the pendulum-attractor separation distance was $d = 50\rm\ \mu m$.  For the Yukawa correction, the parameters $\alpha = 10^{-3}$ and $\lambda = 1\rm\ mm$ were used and, for the power-law correction, we used $k = 1\rm\ \mu m$.  We note that, for the Yukawa correction, the selected combination of $\alpha$ and $\lambda$ falls just outside the parameter exclusion region set by the current state-of-the-art experimental searches for deviations away from the Newtonian ISL~\cite{Tan:2020, Lee:2020}. 

Figure~\ref{fig:torqueOsc} shows the torque oscillations for three values of $N$ between $18$ and $34$ when considering a Yukawa-corrected anisotropic gravitational force.  Although not shown, the results for the power-law correction were qualitatively and quantitatively similar.  The observations summarized below apply to both cases.  The $N \lesssim 20$ curves are dominated by the isotropic torques such that the low-frequency anisotropic component of the signal cannot be resolved.  The isotropic component, on the other hand, exhibits a strong oscillation with a period of $2\pi/\left[N(N-1)\right]$.  Figure~\ref{fig:torqueOsc}(a) shows the $N = 18$ case.  At larger values of $N$, the anisotropic oscillations (period $2\pi/N$) are clearly resolved.  For example, Fig.~\ref{fig:torqueOsc}(b) shows the $N = 24$ case for which the isotropic and anisotropic components of the torque oscillation are comparable leading to a superposition of the two signals.  For $N\gtrsim 30$, the isotropic component is strongly suppressed such that the torque oscillation is completely dominated by the anisotropic component.  The $N = 34$ case is shown in Fig.~\ref{fig:torqueOsc}(c).  

A plot of the isotropic and anisotropic oscillation amplitudes as a function of $N$ are shown in Fig.~\ref{fig:AmpVsN}.  For both the Yukawa and power-law cases, increasing $N$ from $10$ to $30$ suppresses the fundamental isotropic peak by eight orders of magnitude.  The red squares in Fig.~\ref{fig:AmpVsN}, represent the oscillation amplitudes due to the isotropic component of the gravitational force for both Yukawa and power-law corrections.  Because the corrections to Newtonian gravity are small, the isotropic component of the torque oscillations are largely independent of the details of the correction term.

On the other hand, the amplitude of the anisotropic signal has a much weaker dependence on $N$.  For \mbox{$20\le N\le 100$}, a much wider range of values, the amplitude of the anisotropic torque oscillation only changes by a factor of ten.  In Fig.~\ref{fig:AmpVsN}, the green circles and cyan triangles represent the amplitudes of the anisotropic oscillations for Yukawa- and power-law-corrections, respectively.  The two datasets follow nearly identical trends as $N$ is increased.  The near-coincidence between the numerical values of the observed Yukawa and power law amplitudes is merely due to the choice of parameters for the correction terms ($\alpha=10^{-3}$ and $\lambda = 1\rm\ mm$ for the Yukawa correction and $k = 1\rm\ \mu m$ for the power-law correction).  In summary, this analysis of the differential pendulum shows that it is always possible to select a value of $N$ such that the net torque is completely dominated by the anisotropic contribution.  The required value of $N$ and the strength of the anisotropic signal depends on the geometry of the torsion pendulum and the parameters characterizing the gravitational anisotropy.

Finally, we examined the dependence of the differential pendulum's torque oscillations on the size of the $x\times x$ patch grid.  This analysis was similar to that shown in Fig.~\ref{fig:torque}(b) of Sec.~\ref{sec:isotropic} for the conventional geometry with $N$ pendulum and $N$ attractor fins.  Here, we briefly summarize the key findings for the differential pendulum.  Additional details are presented in the Supplemental Material.  

As expected, the amplitudes of both the isotropic and anisotropic torque oscillations asymptotically approach stable values as $x$ is increased.  However, for the isotropic component, the approach to stability slowed as  $N$ was increased.  For example, reducing the patch target mass from $2.5$ to $1.25\rm\ mg$ caused the isotropic oscillation amplitude to decrease by $1.7$ and $36\%$ when $N = 10$ and $24$, respectively.  Therefore, the red squares in Fig.~\ref{fig:AmpVsN} represent an upper limit on the amplitudes of the isotropic torque oscillations.  For the $N=24$, a target patch mass of $1.25\rm\ mg$ resulted in $x = 63$ ($3969$ patches per fin) and a computation time greater than $50\rm\ hours$ for each orientation $\phi$ of the attractor.  Because it is clear that the isotropic component of the signal is strongly suppressed as $N$ is increased, we did not attempt to reduce the target patch mass any further.  

The anisotropic oscillation amplitude approached stability more quickly as $x$ was increased.  Reducing the patch target mass from $2.5$ to $1.25\rm\ mg$ caused the calculated amplitude to decrease by only $0.2$ and $6.6\%$ for the $N = 20$ and $100$ cases, respectively.  Therefore, the green triangles and cyan circles in Fig.~\ref{fig:AmpVsN} accurately represent the expected amplitudes of the anisotropic torques for our chosen differential pendulum geometries and anisotropy parameters. 

\section{Discussion}

It is important to emphasize that, despite having very different dependencies on $r_{ij}^{\left(kl\right)}$, the Yukawa and power-law corrections produce torques that evolve in nearly identical manners as the value of $N$ is changed.  This is a crucial observation because, in most cases, modifications to general relativity are expected to introduce corrections to Newtonian gravity that are linear combinations of the Yukawa- and power-law-corrections that we have considered.  Therefore, if short-range gravitational anisotropies do exist in nature, regardless of the underlying theory, the proposed apparatus will be a sensitive probe of the resulting anisotropic forces. 

The current state-of-the-art torsion pendulum experiments measure toques with a resolution of \mbox{$\approx 25\times 10^{-18}\rm\ N\,m$}~\cite{Lee:2020, Tan:2020}. As shown in Figs.~\ref{fig:torqueOsc} and \ref{fig:AmpVsN}, the anisotropic component of the torque signal has an amplitude between $10^{-21}$ and $10^{-20}\rm\ N\,m$.  Although these calculated torques are below the current experimental sensitivities, it is important to note that the actual strength of the anisotropic signal depends on the both the anisotropy parameter $\varepsilon$ (see Eq. (S5) in the Supplemental Material) and the parameters characterizing the corrections to Newtonian gravity ($\alpha$ and $\lambda$ for the Yukawa correction and $k$ for the power-law correction).  Furthermore, the geometry of the pendulum and attractor can be changed in order to enhance the strength of the anisotropic component of the torque.

Finally, we point out that Zhu {\it et al.}\ have recently described and demonstrated a very sensitive displacement probe with sub-nanometer or picometer resolution~\cite{Zhu:2022}.  The probe is made from a section of open-ended coaxial cable operated as a microwave Fabry-Perot interferometer.  Incorporating such a high-resolution device to track changes in the position of the suspended pendulum could substantially enhance the sensitivity of experimental studies of the gravitational field using torsion pendulums~\cite{Zhu:2022}.

\section{Conclusion}

Motivated by theoretical and empirical arguments for short-distance anisotropy, and the absence of a suitable experiment to detect such anisotropy, we have proposed an apparatus based on a simple, but novel, modification to the torsion-pendulum experiments.  The proposed apparatus has the capability to probe anisotropies at micrometer length scales. We have incorporated anisotropy using both Yukawa-type and power-law-type short-distance corrections to gravity, and observed that the resulting signal is largely independent of the details of the underlying short-distance modification to gravity. Thus, any form of short-distance anisotropic correction can be detected using the proposed apparatus.  An important aspect of this proposal is that it only requires a small modification to proven experimental methods that are already used to search for short-range deviations from the ISL. Thus, the differential torsion pendulum described in this paper can be constructed relatively easily and used to either verify the existence of short-distance anisotropy, or at least set a strong bound on the scale at which such anisotropies can occur. It would also be interesting to use the insights presented in this work to develop experiments to search for anisotropies in other theories, such as those that may arise from short-range corrections to quantum mechanics~\cite{mir}. 

\section*{Acknowledements}
The authors thank Eric G.\ Adelberger (University of Washington) for helpful discussions that improved the manuscript.  We also acknowledge the computational support provided by the Digital Research Alliance of Canada.


\clearpage
\onecolumn

\setcounter{section}{0}
\setcounter{equation}{0}
\setcounter{figure}{0}
\setcounter{table}{0}
\setcounter{page}{1}

\renewcommand{\theequation}{S\arabic{equation}}
\renewcommand{\thesection}{S\arabic{section}}
\renewcommand{\thefigure}{S\arabic{figure}}

\title[Detecting Short-Range Anisotropic Corrections to Gravity: Supplemental Material]{Novel Setup for Detecting Short-Range Anisotropic Corrections to Gravity: Supplemental Material}

\author{Jake S. Bobowski$^1$,  Hrishikesh Patel$^{2,3}$, Mir Faizal$^{3}$} 
\address{$^1$Irving K. Barber Faculty of Science, University of British Columbia, 3333 University Way, Kelowna, British Columbia, Canada V1V 1V7} 
\address{$^2$Department of Physics and Astronomy, University of British Columbia, 6224 Agricultural Road, Vancouver, British Columbia, Canada V6T 1Z1}
\address{$^3$Canadian Quantum Research Center, 204-3002 32nd Avenue, Vernon, British Columbia, Canada V1T 2L7}
\ead{jake.bobowski@ubc.ca}

\begin{abstract}
This Supplemental Material provides additional details about the numerical calculations used to characterize the proposed differential torsion pendulum.  First, we present the geometry of pendulum and attractor fins and relate their positions to the calculated torque.  Second, we discuss how anisotropy was incorporated into the short-range corrections to Newtonian gravity.  Next, the dependence of the anisotropic torque on the angular size of the pendulum and attractor fins is presented.  Finally, we comment on the accuracy of the numerically-calculated isotropic and anisotropic torques.




\end{abstract}


\section{Differential Torsion Pendulum and Anisotropic Torque}
\begin{figure}[b]
    \centering
    \includegraphics[width=0.5\columnwidth]{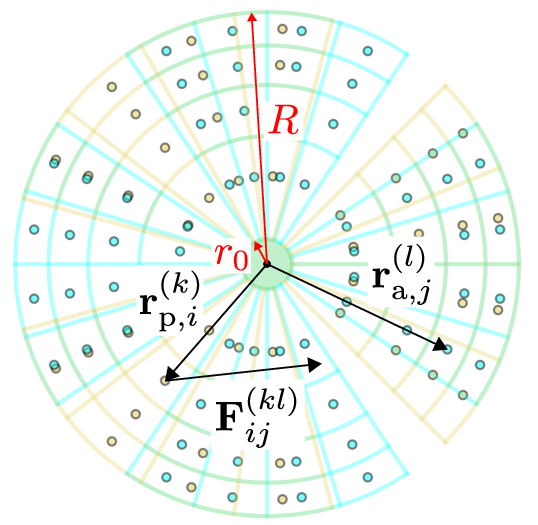}
    \caption{Top view of a pendulum with $N - 1 = 3$ fins suspended above an attractor with $N=4$ fins.  The vectors $\mathbf{r}_{\mathrm{p},i}^{\left(k\right)}$ and $\mathbf{r}_{\mathrm{a},j}^{\left(l\right)}$ locate the center-of-mass positions of patch $k$ in pendulum fin $i$ and patch $l$ in attractor fin $j$, respectively.}
    \label{fig:S1}
\end{figure}

Figure~\ref{fig:S1} shows a top view of a differential torison pendulum with $N - 1 = 3$ pendulum fins and $N = 4$ attractor fins with every fin subdivided into a grid of $x\times x$ patches, where $x = 4$.  The patch boundaries have been chosen such that all patches have the same area.
As shown in the figure, the fins have inner radii $r_0$ and outer radii $R$.  For both the pendulum and attractor, the fins fill $75\%$ of the available area, corresponding to a filling fraction of $f = 0.75$.  In general, the angular widths $\beta_\mathrm{p/a}$ of the pendulum and attractor fins are given by  
\begin{align}
\beta_\mathrm{p} &= \frac{2\pi f}{N - 1}\label{eq:betap}\\
\beta_\mathrm{a} &= \frac{2\pi f}{N}.\label{eq:betaa}
\end{align}
Relative to the center of a fin, the angular position $\delta_\mathrm{p/a}^{\left(m\right)}$ of the center-of-mass of patch $m$ is given by
\begin{equation}
\delta_\mathrm{p/a}^{\left(m\right)} = \frac{\beta_\mathrm{p/a}}{2}\left(\frac{2m + 1}{x} - 1\right),
\end{equation}
where $m = 0, 1, \dots, x - 1$ is an integer.

The radial boundaries of the patches are given by
\begin{equation}
b_n = \sqrt{\left(\frac{n}{x}\right)R^2 + \left(1 - \frac{n}{x}\right)r_0^2},
\end{equation}
where $n = 0, 1, \dots, x$ is another integer. The center-of-mass positions of patch $m$ along the radial direction, measured from the rotation axis, is then given by
\begin{equation}
r^{\left(m\right)} = \frac{2}{3}\left(\frac{b_{m+1}^3 - b_m^3}{b_{m+1}^2 - b_m^2}\right).
\end{equation}
Finally, the centre-of-mass position of patch $k$ in pendulum fin $i$ is given by
\begin{equation}
    \frac{\mathbf{r}_{\mathrm{p},i}^{\left(k\right)}}{r^{\left(k\right)}} =\cos\left(\theta + \delta_\mathrm{p}^{\left(k\right)} + \frac{2\pi i}{N-1}\right)\,\hat\imath + \sin\left(\theta + \delta_\mathrm{p}^{\left(k\right)} + \frac{2\pi i}{N-1}\right)\,\hat\jmath\label{eq:CoMp}
\end{equation}
and the centre-of-mass position of patch $l$ in attractor fin $j$ is given by  
\begin{equation}
    \frac{\mathbf{r}_{\mathrm{a},j}}{r^{\left(l\right)}} =\cos\left(\phi + \delta_\mathrm{a}^{\left(l\right)} + \frac{2\pi j}{N}\right)\,\hat\imath + \sin\left(\phi + \delta_\mathrm{a}^{\left(l\right)} + \frac{2\pi j}{N}\right)\,\hat\jmath -z\,\hat{k}.\label{eq:CoMa}
\end{equation}
Equations (\ref{eq:CoMp}) and (\ref{eq:CoMa}) assume that the centre of mass of each pendulum fin lies in the $xy$-plane ($z=0$) and the centre of mass positions of the attractor fins are a distance $z$ below the $xy$-plane.  If the pendulum and attractor fins each have a thickness $t$ and the top of the attractor fins are a distance $d$ below the bottom of the pendulum fins, then $z=-(t+d)$.  The angles $\theta$ and $\phi$ give the orientations of the pendulum and attractor, respectively, relative to a reference $x$-axis.  As discussed in the main manuscript, the force of patch $l$ in attractor fin $j$ on patch $k$ in pendulum fin $i$ due to Newtonian gravity is given by
\begin{equation}
\mathbf{F}_{ij}^{\left(kl\right)}= \frac{G m_\mathrm{p}m_\mathrm{a}}{\left(r_{ij}^{\left(kl\right)}\right)^2}\,\hat{r}_{ij}^{\left(kl\right)}, 
\end{equation}
where $m_\mathrm{p}$ and $m_\mathrm{a}$ are the masses of the pendulum and attractor patches, respectively, $\mathbf{r}_{ij}^{\left(kl\right)} = \mathbf{r}_{\mathrm{a},j}^{\left(l\right)} - \mathbf{r}_{\mathrm{p},i}^{\left(k\right)}$, and $\hat{r}_{ij}^{\left(kl\right)}$ is a unit vector in the direction of $\mathbf{r}_{ij}^{\left(kl\right)}$.

The isotropic form of the Yukawa-corrected gravitational potential is given by~\cite{Floratos:1999}
\begin{equation}
V_{ij}^{\left(kl\right)} = -\frac{G m_\mathrm{p} m_\mathrm{a}}{r_{ij}^{\left(kl\right)}} \left(1 + \alpha  e^{-r_{ij}^{\left(kl\right)}/\lambda} \right).
\end{equation}
Here, $\alpha$ is the strength correction parameter and $\lambda$ is the scale correction parameter.  The isotropic form of the Yukawa-corrected gravitational force is then given by  
\begin{equation}
\mathbf{F}_{\mathrm{Y},ij}^{\left(kl\right)} = \frac{Gm_\mathrm{p} m_\mathrm{a}}{\left(r_{ij}^{\left(kl\right)}\right)^2} \Bigg[1 + \alpha \left(1 + \frac{r_{ij}^{\left(kl\right)}}{\lambda}\right)e^{-r_{ij}^{\left(kl\right)}/\lambda} \Bigg]\hat{r}_{ij}^{\left(kl\right)}.
\end{equation}
We only consider gravitational anisotropies applied to the Yukawa term. The reason is that any correction (for example, from some theory of quantum gravity) would be a higher-order correction to the Einstein–Hilbert action.  As the Einstein–Hilbert action preserves isotropy, we expect only the correction terms to break the isotropy of space.  To make the Yukawa term anisotropic, we construct a normalized anisotropic unit vector given by
\begin{equation}
\hat{r}^{\mathrm{a},\left(kl\right)}_{ij}(\varepsilon) = \frac{(1+\varepsilon)\sin \theta^{\prime\left(kl\right)}_{ij} \cos \phi^{\prime\left(kl\right)}_{ij}\,\hat\imath \, + \sin \theta^{\prime\left(kl\right)}_{ij} \sin \phi^{\prime\left(kl\right)}_{ij}\,\hat\jmath + \cos \theta^{\prime\left(kl\right)}_{ij}\,\hat{k}}{\sqrt{1 + \varepsilon(2 + \varepsilon)\sin^2\theta^{\prime\left(kl\right)}_{ij} \cos^2 \phi^{\prime\left(kl\right)}_{ij}}}.
\end{equation}
The angles $\theta^{\prime\left(kl\right)}_{ij}$ and  $\phi^{\prime\left(kl\right)}_{ij}$ specify the direction of $\hat{r}^{\mathrm{a},\left(kl\right)}_{ij}(\varepsilon)$ in spherical coordinates and the parameter $\varepsilon$ determines the strength of the anisotropy.  Note that, $\hat{r}^{\mathrm{a},\left(kl\right)}_{ij}\equiv \hat{r}^{\left(kl\right)}_{ij}$ when $\varepsilon = 0$.  The anisotropic Yukawa-corrected force can then be expressed as 
\begin{equation}
\mathbf{F}^{\mathrm{a},\left(kl\right)}_{\mathrm{Y},ij} = \frac{Gm_\mathrm{p} m_\mathrm{a}}{\left(r_{ij}^{\left(kl\right)}\right)^2}\Bigg[\hat{r}^{\left(kl\right)}_{ij} + \alpha \left(1 + \frac{r_{ij}^{\left(kl\right)}}{\lambda}\right)e^{-r_{ij}^{\left(kl\right)}/\lambda}\,\hat{r}^{\mathrm{a},\left(kl\right)}_{ij}(\varepsilon) \Bigg].\label{eq:FY}
\end{equation}
In a similar way, an anisotropic power-law corrected force can be expressed as
\begin{equation}
    \mathbf{F}^{\mathrm{a},\left(kl\right)}_{\mathrm{P},ij} = \frac{G m_\mathrm{p} m_\mathrm{a}}{\left(r_{ij}^{\left(kl\right)}\right)^2} \left[ \hat{r}_{ij}^{\left(kl\right)} + \left(\frac{k}{r_{ij}^{\left(kl\right)}}\right)\,\hat{r}^{\mathrm{a},\left(kl\right)}_{ij}(\varepsilon)\right].\label{eq:FP}
\end{equation}
Equations~(\ref{eq:FY}) and (\ref{eq:FP}) reproduce Eqs.~(2) and (3) of the main manuscript, respectively.

\section{Anisotropic Torque Amplitude Versus Filling Fraction}\label{sec:filling}

All of the results presented in Secs.~\ref{sec:filling} and \ref{sec:patches} assume an anisotropic Yukawa-corrected gravitational force with $\alpha = 1\times 10^{-3}$, $\lambda = 1\rm\ mm$, and $\varepsilon = 1\times 10^{-4}$.  Furthermore, the pendulum and attractor fins were assumed to be made of tungsten ($\rho = 19.3\rm\ g/cm^3$) and to have a thickness of $t = 1\rm\ mm$ and inner and outer radii of $r_0 = 5\rm\ mm$ and $R=50\rm\ mm$, respectively.  The distance separating the bottom of the pendulum and the top of the attractor was set to $d = 50\rm\ \mu m$.  All of these parameter values are consistent with those used in the main manuscript.

\begin{figure}
    \centering
    \includegraphics[width = 0.65\columnwidth]{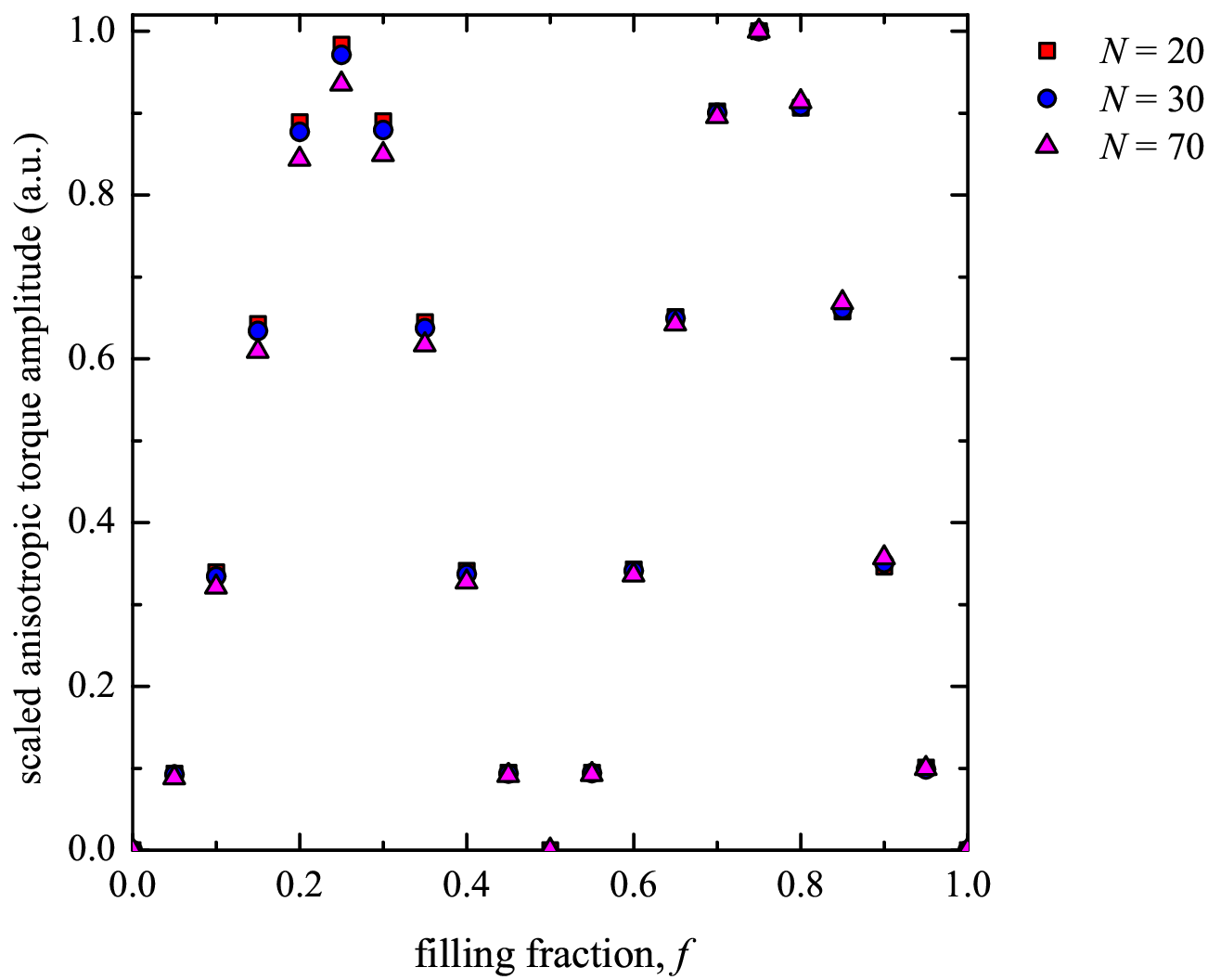}
    \caption{Amplitude of the anisotropic component of the torque oscillation as a function of the pendulum and attractor fin filling fraction $f$.  To facilitate easy comparisons, the data were scaled to make the peak height at $f = 0.75$ equal to one.  Data are shown for $N = 20$, $30$, and $70$.}
    \label{fig:ampVsfilling}
\end{figure}

The first differential torsion pendulum geometry we considered used a fin filling fraction $f = 0.5$ and reveled a vanishingly-small amplitude for the anisotropic component of the torque oscillation.  This observation led us to investigate the dependence of the anisotropic amplitude on the filling fraction of the fins for $0\le f\le 1$.  Figure~\ref{fig:ampVsfilling} shows the results for differential pendulums with $N = 20$, $30$, and $70$.  

As expected, the anisotropic torque vanishes for $f = 0$ (no pendulum and attractor fins) and $f = 1$ (no empty space between neighboring fins).  There is also a zero in the anisotropic torque amplitude at $f = 0.5$, consistent with our original observation.   The oscillation amplitude peaks between these minima at $f = 0.25$ and $0.75$.  In order to compare the results from the different values of $N$, the data in the figure were scaled to make the peak amplitudes at $f = 0.75$ equal to one.  When scaled in this way, Fig.~\ref{fig:ampVsfilling} shows that oscillation amplitude's dependence on $f$ is robust against changes to $N$.  

Notice that the peak at $0.75$ filling is slightly higher than that at $f = 0.25$.  One possible explanation is that the pendulum and attractor fins have higher total mass at the larger filling fraction which leads to greater torque amplitudes.  Another possibility is that the difference in peak heights is due to changes in the patch mass as the fin size or, equivalently, the filling fraction is varied.  For a given value of $N$, all of the points in Fig.~\ref{fig:ampVsfilling} were calculated using the same $x\times x$ patch grid.  For example, the $N = 70$ data were calculated using $x = 10$.  Since the angular widths of the pendulum and attractor fins are proportional to $f$, the fins at $f=0.75$ are three times the size of the fins $f = 0.25$.  As a result, for fixed $x$, the patch masses are smaller for $f = 0.25$ than they are for $f=0.75$.  Intuitively, the smaller patch size is expected to lead to more accurate numerical results.  As will be shown in Sec.~\ref{sec:patches}, the amplitude of the torque oscillations decreases as the accuracy of the numerical calculations is increased.  Therefore, the difference in peak heights at $f = 0.25$ and $0.75$ could be due to the increased numerical accuracy of the calculations at lower filling fractions.

\section{Isotropic and Anisotropic Torque Amplitude Versus Number of Patches}\label{sec:patches}
This section is used to investigate the numerical accuracy of the torque calculations as a function of $x$ for the differential pendulum.  The calculations where done using the parameters given in Sec.~\ref{sec:filling} and a filling fraction of $f = 0.75$.  The analysis is similar to that presented in the main manuscript for the conventional torsion pendulum with $N$ pendulum fins and $N$ attractor fins.

\begin{figure}
    \centering
    \begin{tabular}{c}
    (a)\includegraphics[width = 0.55\columnwidth]{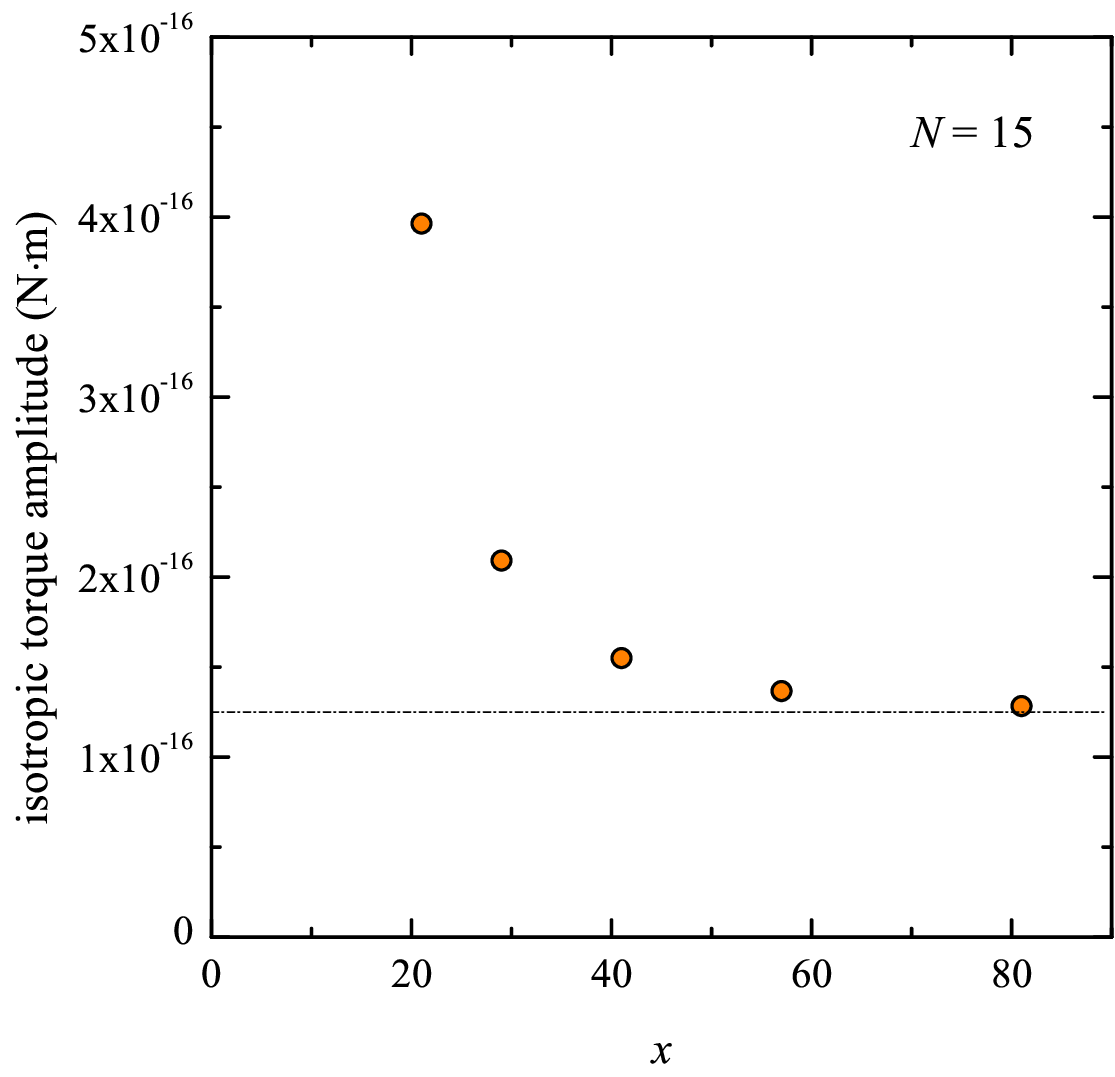}\\
    ~\\
    (b)\includegraphics[width = 0.55\columnwidth]{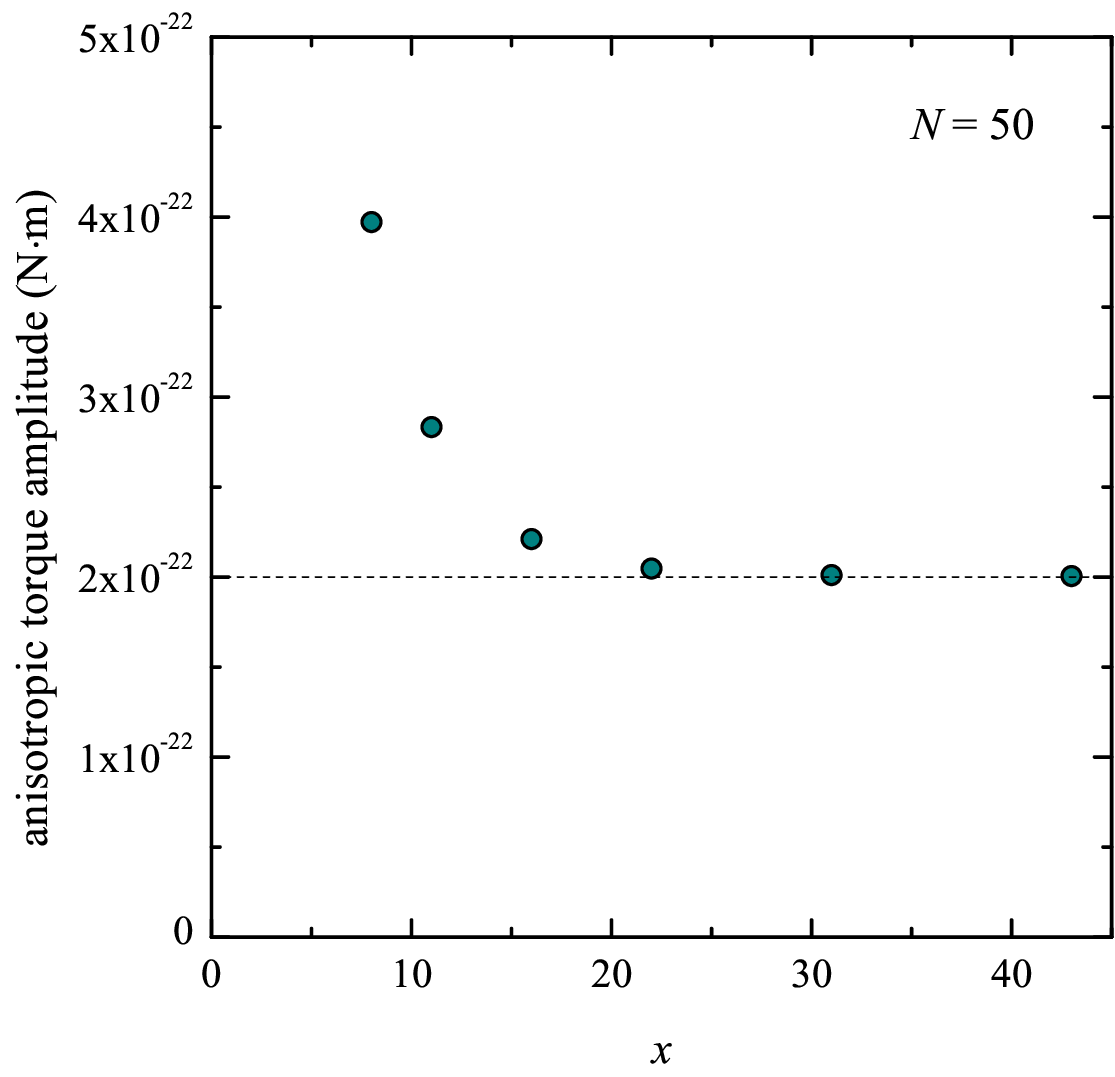}
    \end{tabular}
    \caption{(a) Amplitude of the isotropic component of the torque oscillation as a function of $x$ for $N = 15$.  (b) Amplitude of the anisotropic component of the torque oscillation as a function of $x$ for $N = 50$.  The dashed horizontal lines are guides to the eye.}
    \label{fig:ampVSx}
\end{figure}

Figure~\ref{fig:ampVSx} shows the amplitudes of the isotropic and anisotropic torque oscillations as a function of $x$.  At the maximum value of $x$ used, the pendulum patch masses were less than or equal to $1.25\rm\ mg$.  The attractor fins were divided into $x\times x$ patch grids using the same $x$ selected for the pendulum fins.  Figure~\ref{fig:ampVSx}(a) shows the isotropic torque amplitude versus $x$ for a differential pendulum with $N = 15$.  Even at $x=81$ ($6561$ patches), the amplitude still shows an appreciable decrease as $x$ increases.  Increasing $x$ from $57$ to $81$ (a doubling of the patch mass), causes the oscillation amplitude to decrease by $6.5\%$.  The horizontal dashed line in the figure is a guide to the eye intended to highlight that the amplitude has not yet fully reached a stable value.  The approach to stability slows even further as $N$ is increased.  In Fig.~6 of the main manuscript, $N=34$ is the maximum value for which the isotropic oscillation amplitude is shown.  Its percent decrease when the target pendulum patch mass is decreased from $2.5$ to $1.25\rm\ mg$ was $63\%$.   Because of the prohibitive computation time, we did not attempt to reduce the target patch mass any further.

In contrast, the anisotropic oscillation amplitudes approached stability more quickly, even when considering large values of $N$.  Figure~\ref{fig:ampVSx}(b) shows the $N=50$ case using the same target patch masses used in part (a) of the figure.  As shown by the dashed line, there is minimal change ($2\%$) in the calculated amplitude as $x$ is increased from $31$ to $43$.  Even when $N$ is increased to $100$, the change in the anisotropic oscillation amplitude is less then $7\%$ when the patch mass is reduced from $2.5$ to $1.25\rm\ mg$. 

\end{document}